\documentclass[11pt,a4paper]{article}
\usepackage{jheppub}

\usepackage{pstricks}
\usepackage{color}
\usepackage{amsmath}

\usepackage{bbm}
\usepackage{verbatim}   % useful for program listings
\usepackage{subfigure}  % use for side-by-side figures
\usepackage{acronym}

\newcommand{\sref}[1]{section~\ref{#1}}
\newcommand{\aref}[1]{appendix~\ref{#1}}

\newcommand{\tref}[1]{table~\ref{#1}}
\newcommand{\eref}[1]{equation~\eqref{#1}}
\newcommand{\mb}[1]{\boldsymbol{#1}}

\newcommand{\kahler}{K\"{a}hler }
\def\Tr{\mathop{\rm Tr}}

\newcommand{\rep}[1]{\mathbf{#1}}
\newcommand{\conjrep}[1]{\overline{\mathbf{#1}}}

\renewcommand{\a}{\alpha}

\newcommand{\e}{\epsilon}
\newcommand{\D}{\Delta}
\renewcommand{\l}{\lambda}
\renewcommand{\th}{\theta}
\newcommand{\IZ}{\mathbb{Z}}

\newcommand{\cL}{\mathcal{L}}
\newcommand{\cR}{\mathcal{R}}

\newcommand{\GeV}{\,\text{GeV}}

\begin{document}

\title{A Supersymmetric One Higgs Doublet Model}

\author[a]{Rhys Davies,}
\emailAdd{daviesr@maths.ox.ac.uk}

\author[b]{John March-Russell,}
\emailAdd{jmr@thphys.ox.ac.uk}

\author[b]{Matthew McCullough}
\emailAdd{mccull@thphys.ox.ac.uk}

%\preprint{OUTP-11-36P}

\affiliation[a]{Mathematical Institute,
University of Oxford\\
24-29 St Giles, Oxford,
OX1 3LB, UK}

\affiliation[b]{Rudolf Peierls Centre for Theoretical Physics,
University of Oxford\\
1 Keble Road, Oxford,
OX1 3NP, UK}

\date{\today}

\abstract{
We present a supersymmetric extension of the Standard Model in which only one
electroweak doublet acquires a vacuum expectation value and gives mass to Standard
Model fermions.  As well as the novel accommodation of a Standard Model Higgs within
a supersymmetric framework, this leads to a very predictive model, with some
advantages over the MSSM.  In particular, problems with proton decay, flavour changing
neutral currents and large $CP$ violation are ameliorated, primarily due to the presence
of an anomaly-free $R$-symmetry.  Since supersymmetry must be broken at a low scale,
gravity-mediated effects which break the $R$-symmetry are naturally small.  The
$R$-symmetry requires the presence of adjoint chiral superfields, to give Dirac
masses to the gauginos; these adjoints are the only non-MSSM fields in the visible
sector.  The LSP is a very light neutralino, which is mostly bino.
Such a light neutralino is not in conflict with experiment, and is a striking prediction of
the minimal model.  Additional scenarios to raise the mass of this neutralino to the
weak scale are also outlined.  Prospects for discovery at the LHC are briefly discussed, along with viable scenarios for achieving gauge-coupling unification.
}

\maketitle

\section{Introduction}\label{intro}

It is common lore that a supersymmetric extension of the Standard Model (SM) requires
the existence of two Higgs doublets, $H_u$ and $H_d$, in order to give masses to both
up-type and down-type quarks (as well as charged leptons).  This is because
holomorphy forbids the usual SM down-type Yukawa couplings
$H_u^\dagger Q D^c$.  Furthermore, the fermionic partners of $H_u$ also contribute to gauge
anomalies, which are cancelled by the partners of $H_d$.  Here we pursue the goal of simplifying the Higgs structure of supersymmetric extensions of
the SM, as opposed to commonly studied variants of the MSSM which often involve a more
complex Higgs sector.

It has been noticed by many authors that after supersymmetry (SUSY) breaking there is
typically a contribution to down-quark masses from induced couplings to $H_u$
(e.g. \cite{Ross:1985by,Babu:1998bf,Ross:2007az,Dine:2007xi,Nandi:2008zw,Graham:2009gr,Dobrescu:2010mk}).  Recently,
it was suggested that down-type masses might arise solely from such supersymmetry-breaking
couplings, so that either $H_d$ acquires a vacuum expectation value (VEV) but does not
couple to SM fermions, or can be left out of the spectrum entirely \cite{Graham:2009gr,Dobrescu:2010mk,Ibe:2010ig}.  The latter
necessitates the introduction of a number of new fields in the electroweak sector, in order to
cancel gauge anomalies.

In this paper, we advocate a different scenario.  As the most economical way to cancel
the gauge anomalies of $H_u$, the doublet $H_d$ is retained, but forbidden from
acquiring a VEV, or coupling directly to SM fermions.  As such, it is not really a `Higgs' field
at all; to reflect this we re-label it as $\eta$, and the usual $H_u$ field simply as $H$.
In fact, as the usual $B\mu$-induced mixing with $\eta$ is forbidden, the bosonic components
of $H$ become indistinguishable from a SM Higgs.  We will refer to this setup
as the `Supersymmetric One Higgs Doublet Model' (SOHDM).

A very natural way to implement the above scenario is to impose an anomaly-free $R$-symmetry,\footnote{Although we assume a $U(1)$ $R$-symmetry throughout, the $\IZ_p$ subgroup for sufficiently large $p$ does the
same job and is well motivated from a theoretical perspective \cite{Dine:2009swa}.  Previously discussed models which consider some semblance of an $R$-symmetry, whether in the full model or solely in the gauge sector include \cite{Fayet:1978qc,Polchinski:1982an,Hall:1990hq,Kurosawa:2001iq,Fox:2002bu,Antoniadis:2005em,Antoniadis:2006eb,Antoniadis:2006uj,Amigo:2008rc,Blechman:2009if,Benakli:2008pg,Belanger:2009wf,Benakli:2009mk,Chun:2009zx,Benakli:2010gi,Carpenter:2010as,Kribs:2010md,Lee:2010gv,Abel:2011dc,Lee:2011dy}.} which we describe in \sref{sec:model}.  We stress that this is a different
$R$-symmetry to those commonly studied in the literature, and has a number of
comparative advantages.  In particular, the field content of the model is smaller than
that of the `Minimal $R$-symmetric Supersymmetric Standard Model' (MRSSM) of \cite{Kribs:2007ac}.
The $R$-symmetry forbids Majorana gaugino masses, so we are required to add chiral
superfields in the adjoint of $SU(3){\times}SU(2)$, in order to give the gauginos Dirac
masses.  These are the only fields in our model which do not appear in the MSSM.
The $R$-symmetry allows a $\mu$-term, so there is no need to introduce the extra
doublets of the MRSSM to generate acceptable chargino masses.

We expect effects from Planck-scale physics to violate any global symmetry, (see e.g.\ \cite{Giddings:1987cg,Giddings:1988wv,Giddings:1989bq,Coleman:1989zu,Abbott:1989jw,Barr:1992qq,Kamionkowski:1992mf,Holman:1992us,Kallosh:1995hi,Banks:2010zn}).  Henceforth when referring to the $R$-symmetry we do so in the understanding that it is broken either by Planck-suppressed, or non-perturbatively small operators.  As such our continuous $R$-symmetry is to be viewed as an emergent `accidental' symmetry of the low-energy theory.

The following sections will describe the model in detail, but it may be useful to
list its main features here:
\begin{itemize}
    \item  Couplings between the Higgs and SM fermions are the same as for the SM Higgs.
    \item  One-loop corrections to the Higgs mass from stop/top loops automatically saturate the MSSM correction, reducing fine-tuning.
    \item  Supersymmetry must be broken at a low scale, in order to generate acceptable
    down-type quark masses.  This suggests a gauge-mediated scenario, but the
    usual $\mu/B\mu$ problem of such models is avoided, since the $B\mu$ term
    is neither present nor required.
    \item  $R$-symmetry should be preserved at the TeV scale, allowing simple models of SUSY-breaking
    to be employed, as comparable SUSY and $R$-symmetry breaking are not required.  $R$-symmetry
    breaking effects arising from the expectation value of the superpotential, which is required in order to set the vacuum energy density to a small value, are negligible.
    \item  The anomaly-free $R$-symmetry forbids all dimension four and five baryon number
    violation, implying a proton lifetime well above the experimental lower bound.  This is in contradistinction to the MSSM, where dangerous dimension five operators are allowed by R-parity and must be suppressed by some other means.
    \item  Flavour physics is necessarily connected to supersymmetry breaking, in contrast to standard assumptions.
    \item  The $R$-symmetry has profound implications for flavour physics and $CP$-violation.  All $A$-terms are forbidden, as are Majorana masses for the gauginos, and these two facts significantly reduce the contributions to flavour-changing neutral currents (FCNCs) and $CP$-violation relative to the MSSM \cite{Kribs:2007ac}.
    \item  The symmetries of the model allow small neutrino masses to be generated via
    a standard high-scale seesaw mechanism.
    \item  The field content differs from that of the MSSM only in the addition of chiral superfields
    in the adjoint of $SU(3){\times}SU(2)$, required to give gauginos mass in the presence of
    unbroken $R$-symmetry.
    \item  The minimal version of the model predicts a very light neutralino, which
    is nevertheless consistent with all experimental constraints.
\end{itemize}

The layout of the rest of the paper is as follows.  In \sref{sec:model} we describe the field
content and symmetries of the model and in \sref{sec:massscales} we describe the mass spectrum, including the generation of small neutrino masses.  Between them, \sref{sec:model} and \sref{sec:massscales} contain all of the original details of the SOHDM, and are thus a self-contained reference for a reader interested purely in the structural features of the model.  Following these sections we discuss constraints and phenomenological aspects, and review work by previous authors.  Constraints from flavour physics,
$CP$-violation and precision electroweak physics are discussed in \sref{sec:R-symmetry}.
A brief discussion of gauge coupling unification is included in \sref{sec:couplings}.  Potential
collider signals and ways to unambiguously distinguish this model from other supersymmetric models are discussed in \sref{sec:collider}.  We conclude in \sref{sec:con}.

Finally a word on notation.  We will write superfields in bold, and use the same symbol for their individual components.
Component fields with $R$-charge $\pm 1$ (squarks, sleptons, `--inos') will carry tildes.  So
the left-handed electron superfield, for instance, is
\begin{equation}
    \mb{e_L} = \tilde{e}_L + \sqrt{2}\,\th\, e_L + \ldots~.
\end{equation}
As already mentioned, since there is only one Higgs doublet, we refer to the superfield with the quantum numbers of the up-type Higgs as $\mb{H}$, and the superfield with the quantum numbers of the usual down-type Higgs as $\mb{\eta}$.

\section{The model}\label{sec:model}  %% Matthew

The field content of the model is that of the MSSM, along with chiral superfields in the
adjoint of $SU(3){\times}SU(2)$, needed to give Dirac masses to the gauginos.\footnote{See
below for why we omit the $U(1)_Y$ adjoint i.e.\ a singlet.}  We
will denote by $\mb{O}$ the $SU(3)$ octet, and by $\mb{T}$ the $SU(2)$ triplet.  We summarise
the spectrum, including gauge and global-symmetry charges, in \tref{tab:spectrum}.
\begin{table}[t]
\begin{center}
%% The following adds some white space between row-dividing lines and text %%
\def\str{\vrule height12pt width0pt depth7pt}
\begin{tabular}{| l | l | c | r | r |}
    \hline\str
    ~Field ~&~ Gauge rep. ~&~ $R$-charge ~&~ $\IZ_2$-parity~ \\
    \hline\str
    ~~$\mb{Q}$ & ~~$(\rep{3},\rep{2}, \frac{1}{6})$ & $1$ & $1$\hspace{2em} \\
    \hline\str
    ~~$\mb{U^c}$ & ~~$(\conjrep{3},\rep{1}, -\frac 23)$ & $1$ & $1$\hspace{2em} \\
    \hline\str
    ~~$\mb{D^c}$ & ~~$(\conjrep{3},\rep{1}, \frac 13)$ & $1$ & $-1$\hspace{2em} \\
    \hline\str
    ~~$\mb{L}$ & ~~$(\rep{1},\rep{2}, -\frac 12)$ & $1$ & $1$\hspace{2em} \\
    \hline\str
    ~~$\mb{E^c}$ & ~~$(\rep{1},\rep{1}, 1)$ & $1$ & $-1$\hspace{2em} \\
    \hline\str
    ~~$\mb{H}$ & ~~$(\rep{1},\rep{2}, \frac 12)$ & $0$ & $1$\hspace{2em} \\
    \hline\str
    ~~$\mb{\eta}$ & ~~$(\rep{1},\rep{2}, -\frac 12)$ & $2$ & $-1$\hspace{2em} \\
    \hline\str
    ~~$\mb{O}$ & ~~$(\rep{8},\rep{1}, 0)$ & $0$ & $1$\hspace{2em} \\
    \hline\str
    ~~$\mb{T}$ & ~~$(\rep{1},\rep{3}, 0)$ & $0$ & $1$\hspace{2em} \\
    \hline\str
    ~~$\mb{X}$ & ~~$(\rep{1},\rep{1},0)$ & $2$ & $-1$\hspace{2em} \\
    \hline\str
    ~~$\mb{W'}$ & ~~$(\rep{1},\rep{1},0)$ & $1$ & $1$\hspace{2em} \\
    \hline
\end{tabular}
\parbox{.65\textwidth}
{\caption{\label{tab:spectrum}
\small The chiral superfield matter content of the SOHDM.  Gauge superfields are even under the $\IZ_2$-parity and are not shown.  The fields $\mb{X}$ and $\mb{W'}$ are the spurion superfields parametrising SUSY breaking.}}
\end{center}
\end{table}

Throughout, SUSY breaking will be parametrised by two spurion chiral superfields.  The
first source is the $F$-term of a chiral superfield $\mb{X}$, to which we assign
$R$-charge $2$ so that $R$-symmetry remains unbroken, and the second is the $D$-term
of an additional $U(1)'$ gauge superfield $\mb{W'}$.  We will denote by $M$ the generic
messenger mass scale, by which we suppress all interactions with $\mb{X}$ and $\mb{W'}$.
Note that the $\mu$-term is not forbidden by our $R$-charge assignments, so in order to
solve the $\mu$-problem, we also impose a discrete $\IZ_2$ symmetry which is broken by
the $F$-term of $\mb{X}$; the corresponding parities are given in \tref{tab:spectrum}.
We will now proceed to discuss the interactions allowed by this structure.

Due to its $R$-charge, $\eta$ has no Yukawa couplings to SM fermions.  Fermion
masses therefore come entirely from Yukawa couplings to $H$, with the charged lepton and
down-type quark couplings induced by supersymmetry breaking:\footnote{For a model that
generates this structure see \cite{Ibe:2010ig}.}
\begin{equation}
   \cL_{\text{Yuk}} =  \int\! d^2\th\,\l_U\,\mb{H} \mb{Q} \mb{U^c} + \int\!d^4\th\, \frac{\mb{X^\dagger H^\dagger}}{M^2}(
            \l_D\,\mb{Q D^c} + \l_E\, \mb{L E^c} )~.
\label{eq:yuk}
\end{equation}
We therefore have a tree-level relation for the bottom-quark mass,
\begin{equation}\label{eq:FMsquared}
    \frac{\l_b F_X}{M^2} 174\GeV \simeq 5\GeV ~~\Rightarrow~~\frac{F_X}{M^2} \simeq \frac{1}{35\l_b}~.
\end{equation}
If we require that the bottom quark coupling
is perturbative, say $\l_b \lesssim 1$, this gives a lower bound:
\begin{equation}
    \frac{F_X}{M^2} \gtrsim \frac{1}{35}~.
\end{equation}

In the Higgs sector, the $\mu$-term is generated by a low-scale Giudice-Masiero
mechanism, and $\mb{H}$ also has a renormalisable superpotential coupling to
$\mb{T}$ and $\mb{\eta}$ generated after SUSY-breaking:
\begin{equation}\label{eq:Higgslag}
    \cL_{\text{Higgs}} = \int\!d^4\th\, \frac{\mb{X^\dagger}}{M} (\l_\mu \mb{H \eta} +\frac{ \l'_T}{M} \,\mb{H T \eta}) \to \int\!d^2\th\, (\mu \mb{H \eta} +\l_T \,\mb{H T \eta})~.
\end{equation}
The effective $\mu$-term is therefore given by $\mu = \l_\mu F_X/M$; combining this
with \eqref{eq:FMsquared}, we get
\begin{equation}
    M \simeq 35\frac{\l_b}{\l_\mu} \mu ~,~~ F_X \simeq 35 \frac{\l_b}{\l_\mu^2}\mu^2~.
\end{equation}
The requirements of a weak-scale $\mu$-term and natural couplings therefore
dictate that the scale of SUSY breaking is low.  Notice also that \eqref{eq:Higgslag}
implies that the $\mb{T}$ Yukawa coupling is small,
$\l_T \lesssim \mathcal{O}(F_X/M^2) \sim 1/35$.

Dirac mass terms for the gauginos and their adjoint partners can be written as
\begin{equation}
        \cL_\text{D} = \int\!d^2\th\, \frac{\mb{W}'_\alpha}{M} \big(\l_G\Tr(\mb{O G^\a}) + \l_W\Tr(\mb{T W}^\a)\big)
            \to M_3\Tr(\widetilde O \widetilde G) + M_2 \Tr(\widetilde T \widetilde W) + \ldots~,
\end{equation}
where $M_3 = \l_G D'/M$, and $M_2 = \l_W D'/M$.

Finally, soft scalar masses are given by K\"ahler terms of the form
\begin{equation}\label{eq:softlag}
    \cL_\text{soft} = \int\!d^4\th \frac{\mb{X^\dagger X}}{M^2} \big(\a_Q^2\mb{Q^\dagger Q} + \ldots\big)
\end{equation}
where after $\mb{Q^\dagger Q}$ we insert analogous terms for all chiral fields, to generate
soft masses for their scalars.  Because the adjoint is a real representation, and our adjoint
chiral fields have $R$-charge zero, there are also holomorphic terms for their scalars, i.e.\
for the triplet we get both $\Tr(\mb{T^\dagger T})$ and  $\Tr(\mb{T}^2)$.  The second of these terms
gives squared masses to the real and imaginary components of $T$ which are of equal
magnitude but opposite sign, but this is not a problem if such terms are sub-dominant.

\subsection{Discussion}

We are assuming a combination of $F$- and $D$-term SUSY breaking in the hidden sector in
order to generate large enough gaugino masses (see \sref{sec:massscales}).\footnote{Given
recent results on $R$-symmetric gauge mediation \cite{Abel:2011dc}, it may in fact be possible
to achieve the same effective softly broken Lagrangian purely from $F$-term SUSY breaking.}
This is not unreasonable, as although dominant $D$-term breaking
does not arise dynamically \cite{Komargodski:2009pc,Dumitrescu:2010ca} it is possible for a
hidden sector to give rise to SUSY breaking satisfying $D \lesssim F$, even for non-Abelian
$D$-terms \cite{Dumitrescu:2010ca}.  Some examples of hidden sectors with mixed $F$- and
$D$-term SUSY breaking can be found in
\cite{Raby:1998bg,Dienes:2008gj,Seiberg:2008qj,Elvang:2009gk,Matos:2009xv,Dumitrescu:2010ca}.

Previous attempts to build models with an $R$-symmetry have required VEVs for both $H_u$
and $H_d$, and can be split into two classes.  In the first, $R$-charges of $\cR_{\mb{H_u}} = 0$,
$\cR_{\mb{H_d}} = 2$ are assigned, and the $R$-symmetry is then broken at the level of a few
GeV \cite{Nelson:2002ca} in order to generate a small down-Higgs VEV and acceptable down-type fermion
masses.  In the second, the $R$-charges are taken to be $\cR_{\mb{H_u}} = \cR_{\mb{H_d}} = 0$,
such that $H_d$ can get a VEV with $R$-symmetry remaining unbroken.  In this case it is
necessary to extend the Higgs sector by adding two extra doublets, as in the MRSSM, in order to
generate acceptable Higgsino masses \cite{Kribs:2007ac}.  The novelty herein is that, as a VEV
for $H_d \equiv \eta$ is no longer required, we can assign it an $R$-charge of 2 and still get weak-scale
chargino masses without the addition of extra $SU(2)$ doublets.  Thus the particle content is more
minimal than that of the MRSSM, but an unbroken $R$-symmetry is maintained.

The extra $\IZ_2$ symmetry has been imposed in order to forbid a tree-level $\mu$-term,
thus solving the $\mu$ problem in the usual way \cite{Giudice:1988yz}, but we will see that it has
other desirable consequences, in particular forbidding too-large neutrino masses.

The reader may wonder why we did not also add a singlet chiral field $\mb{S}$ of $R$-charge
$0$, in order to give the bino a weak-scale mass like the other gauginos.  The reason
is that \kahler potential terms such as $\mb{X^\dagger X S}$ and superpotential terms such as
$\mb{W}'^\alpha \mb{W}'_\alpha \mb{S}$ would be allowed, leading to a large VEV for $S$, which
causes problems for the breaking of both supersymmetry and electroweak symmetry.  As we
will discuss in \sref{sec:binomass}, these troublesome tadpole terms can be avoided by imposing
certain restrictions on the messenger couplings, but we find the simplest solution to this
problem is to simply exclude the singlet from the model.  This leads to the striking prediction of
a very light neutralino, which nevertheless avoids all experimental bounds, as we explain
in \sref{sec:bino}.  In \sref{sec:binomass} we also discuss additional scenarios for making this
neutralino more massive.

%%%
\subsection{Electroweak symmetry breaking}\label{electroweak}  % Rhys

Electroweak symmetry breaking is affected by the presence of the triplet $\mb{T}$ and
the associated Dirac gaugino mass term.  This gives new contributions to the $D$-terms
of the $SU(2)$ gauge fields, which pick up pieces linear in the triplet scalars:
\begin{equation}
    \D D^j = -M_2 (T^j + \overline{T^j})~.
\end{equation}
We will see below that this leads to a small VEV for the neutral component $T^0$,
but for now we will ignore this, a posteriori justifying this approximation.

We will assume that all squared soft masses are positive at the messenger scale,
and appeal to radiative electroweak symmetry breaking.  Indeed, if the top squark masses
are $m_{\widetilde{Q},\widetilde{U}} \gtrsim 700$ GeV at the messenger scale ($\sim 100$ TeV)
then, upon running down to the weak scale, the squared soft mass of $H$ is driven negative
as a result of the large top Yukawa \cite{Dimopoulos:1996yq},\footnote{The squared soft-mass
for $H$ at the weak scale is given approximately by;
\begin{equation}
\tilde{m}_{H}^2 (m_{\tilde{t}}) \simeq {\tilde{m}}_{H}^2 (M) - \frac{3}{8 \pi^2} \l_t^2 (m_{{\tilde{t}}_L}^2 + m_{{\tilde{t}}_R}^2) \log(M/m_{\tilde{t}}) ~,
\end{equation}
where $M$ is the messenger scale.} while all others remain positive.  In the
MSSM, this nevertheless leads to a VEV for $H_d$ due to mixing induced by the $B\mu$ term, but
here this and similar operators are forbidden by the symmetries of the model.  So we may
proceed in the knowledge that only $H$ acquires a VEV due to radiative electroweak
symmetry breaking.

With all other fields set to zero,
the Higgs potential is in fact the same as in the SM, but with the coefficient
of the quartic term determined at tree-level by the gauge couplings (by convention we take
$\tilde m_h^2$ positive):
\begin{equation}
    V_{\text{Higgs}} = \frac 18 (g^2 + g'^2)|H|^4 + (|\mu|^2 - \tilde m_{H}^2)|H|^2~,
\end{equation}
leading to a VEV
\begin{equation} \label{eq:HiggsVEV}
    \langle H^0 \rangle = 2 \sqrt{\frac{\tilde m_{H}^2 - | \mu |^2}{g^2 + g'^2}}
        = \frac{v}{\sqrt{2}} \simeq 174 \text{GeV}~,
\end{equation}
where the last equality is fixed by experiment.

We now have to ask whether this is in fact a stable
vacuum.  The form of the soft masses guarantees that the Hessian of
the potential will be positive definite, so an instability can only manifest as a
non-vanishing linear term.  Since $U(1)_\text{EM}$ remains unbroken, we cannot
have linear terms in any charged fields, leaving only $\eta^0$ and $T^0$. Inspection
of the $F$-terms following from \eqref{eq:Higgslag} is enough to see that there is
no linear term in $\eta^0$ (alternatively, note that $\eta$ is charged under the
unbroken $R$-symmetry, which therefore forbids a linear term).  But with
$\langle H^0 \rangle = v/\sqrt{2} = 174~\text{GeV}$ and all other fields set to zero, the
$F$-term for $\eta^0$ and the $D$-term for the neutral generator of $SU(2)$, along
with the soft mass for $T$, give a potential for $T^0$, the non-constant part of which
is;\footnote{All parameters can be taken real, as we will discuss in \sref{sec:CP}.}
\begin{equation}
    V_T = \frac{v^2}{2}(\l_T \mu - \frac 12 g M_2)(T^0 + \overline{T^0})
        + (m_T^2 + \frac{1}{2}\l_T^2 v^2 + 2M_2^2) |T^0|^2 ~.
\end{equation}
We see that there is a linear term in the real part of $T^0$, and that it acquires a VEV:
\begin{equation}\label{eq:tripletVEV}
    \langle T^0 \rangle = \frac{\left( \frac 12 g M_2 - \l_T \mu \right) v^2}{4M_2^2 + \l_T^2 v^2 + 2 m_T^2}~.
\end{equation}
This naturally comes out at $\lesssim 1 \GeV$ which, as we discuss in
\sref{sec:R-symmetry}, is compatible with precision electroweak measurements.  The
fact that this VEV is much smaller than $\langle H \rangle$ justifies our
perturbative calculation.

%%%
\section{Mass scales}\label{sec:massscales}  % Rhys

Although we are considering an effective theory approach to TeV-scale model building, it is instructive to consider the scales of soft parameters in complete models of $R$-symmetric gauge mediation.  Previous discussions of $R$-symmetric gauge mediation and/or Dirac gauginos in gauge mediation include \cite{Fox:2002bu,Antoniadis:2006uj,Amigo:2008rc,Blechman:2009if,Benakli:2008pg,Benakli:2009mk,Benakli:2010gi,Carpenter:2010as,Abel:2011dc}, however we focus here on the results of \cite{Benakli:2010gi}.

In \cite{Benakli:2010gi} the messengers which mediate SUSY breaking have SM gauge charges, but also couple directly to the extra chiral adjoint fields through Yukawa couplings, which we denote $\l$.  As a result the adjoint soft masses-squared arise at one-loop, along with the Dirac gaugino masses, and sfermion soft masses-squared arise at two loops.  The generic prediction for soft masses in this model is given as \cite{Benakli:2010gi}:
\begin{itemize}
\item  Gaugino masses $\sim \frac{\l g}{16 \pi^2} \frac{D'}{M}$
\item  Sfermion masses $\sim \frac{g^2}{16 \pi^2} \frac{F_X}{M}$
\item  Adjoint scalar masses $\sim \frac{\l}{4 \pi} (\frac{D'}{M},\frac{F_X}{M} )$
\end{itemize}
Hence for $\l \sim g$, gaugino and sfermion masses are roughly equal and the adjoint scalars are more massive by a factor of $\sim 4 \pi / g$.  Thus the spectrum is similar to that of a gauge-mediation scenario, with the additional adjoint scalar masses an order of magnitude greater than the SM superpartners.
\subsection{The Higgs}\label{sec:Higgs}

The Higgs sector in the SOHDM is the same as in the SM (up to a small
mixing with the neutral component of the triplet $T$, which we ignore here and discuss further in \sref{sec:R-symmetry}),
and in particular it is much simpler than in the MSSM.  As usual, we fix the gauge so that
the VEV of $H$ is real and positive.  If we package all contributions to the
low-order Higgs potential into the effective parameters $m_h$ and $\l_h$, we can write
\begin{equation}
    V_{\text{Higgs}} = -\frac{m_h^2}{2}|H|^2 + \frac{\l_h}{4}|H|^4~.
\end{equation}
Here $m_h$ is the mass of the physical Higgs boson, and if we write the VEV as
$\langle H \rangle = v/\sqrt{2}$, we obtain the relation
\begin{equation}
    m_h^2 = \frac{\l_h v^2}{2}~.
\end{equation}
Since $v$ is fixed, this is a relation between $m_h$ and $\l_h$.  To leading order, the
coefficient $\l_h$ is just $\frac 12 (g^2 + g'^2)$,\footnote{New contributions to the $SU(2)$ D-terms involving $T$ lead to small reductions to the quartic coupling of $\mathcal{O} (\frac{\alpha}{4 \pi})$.} and therefore we get
$m_h = M_Z$, the same as the tree-level upper bound in the MSSM.  In addition, the large contributions to the Higgs mass arising from loops involving stop squarks and top quarks saturate the MSSM correction.  Thus the Higgs mass-squared is;
\begin{equation}
    m_h^2 = M_Z^2 + \frac{3}{4 \pi^2} \l_t^2 m_t^2 \log \bigg(\frac{m_{\tilde{t}_L} m_{\tilde{t}_R}}{m_t^2} \bigg) ~,
\end{equation}
whereas the MSSM correction is suppressed by a factor of $\cos^2(\alpha)$, where $\alpha=0$ corresponds to the situation where the Higgs boson lives entirely in the $H_u$ doublet.  This feature is attractive from a fine-tuning perspective as both the tree-level Higgs mass and the one-loop correction are necessarily at the upper bounds of the MSSM values.

Depending on how SUSY breaking is mediated to the visible sector an additional reduction in fine-tuning might also be obtained from the following operator;
\begin{equation}
    \int\! d^4\th \frac{\mb{X^\dagger X}}{M^4}(\mb{H^\dagger H})^2 ~\to~ \frac{F_X^2}{M^4}|H|^4~,
\end{equation}
as this operator increases the Higgs quartic coupling and, if sizable, would lead to a greater Higgs mass.

\subsection{Fermionic superpartners}

The chargino mass matrix is
\begin{align}
    & \hspace{2em} \widetilde T^+ \hspace{1.5em} \widetilde H^+ ~~~ \widetilde W^+ \\
    \begin{array}{r}
        \widetilde W^- \\
        \widetilde {\eta\,}^- \\[.5ex]
        \widetilde T^-
    \end{array} &   
    \left(\begin{array}{c c c}
    M_2 & \sqrt{2} M_W & 0 \\
    -\frac{\l_T v}{\sqrt{2}} & \mu & 0 \\[.5ex]
    0 & 0 & M_2 \end{array}\right)~.
\end{align}
The zero entries are enforced by the unbroken $R$-symmetry.  So there is one charged Dirac
fermion of mass $M_2$, coming from the third rows and columns, while the other two mass
states come from the upper left $2\times 2$ block.

% which is formally the same as in the MSSM with $\tan\b \to \infty$, so the
%squared masses are
%\begin{equation}
%    \frac 12\bigg(M_2^2 + |\mu|^2 + 2M_W^2 \pm \sqrt{\big( (M_2 + |\mu|)^2
%        + 2M_W^2\big)\big( (M_2 - |\mu|)^2 + 2M_W^2\big)}\bigg)
%\end{equation}

The neutralino mass matrix is more interesting.  We have one extra neutralino
compared to the MSSM, coming from $\mb{T}$.  Three of the neutralinos (the wino, the
bino, and the neutral fermion in $\mb{\eta}$) have $R$-charge $1$, while the other two
(the neutral fermions from $\mb{H}$ and $\mb{T}$) have $R$-charge $-1$, so as for the
charginos, all masses are Dirac-type, with mass matrix
\begin{align}
    & \hspace{2.0em} \widetilde \eta^0 \hspace{2.8em} \widetilde W^0 \hspace{3.4em} \widetilde B^0 \\
    \begin{array}{r}
        \widetilde H^0 \\[1ex]
        \widetilde T^0
    \end{array} &
    \left(\begin{array}{r c c}
    \mu~~ &~ -M_Z \cos\th_W~ & M_Z \sin\th_W \\[1ex]
    -\frac{\l_T v}{\sqrt{2}} & M_2 & 0 \end{array}\right)~.
\end{align}
Clearly one linear combination of the $R$-charge $1$ neutralinos remains
massless;\footnote{This prediction is relaxed if we introduce additional fields in order to raise the neutralino mass, as described in \sref{sec:binomass}.} explicitly, it is
\begin{equation}
    \left(-1 ~, -\frac{\l_T v}{M_2\sqrt{2}} ~,~ \frac{(\mu M_2 - \l_T v M_Z \cos\th_W)}{M_2 M_Z\sin\th_W}\right)~.
\end{equation}
Compared to $\mu, M_2$, we have $\l_T v \simeq 0$, so this is approximately
\begin{equation}
    \left(-1~, 0~,~\frac{\mu}{M_Z\sin\th_W}\right)~.
\end{equation}
Since $\mu \gtrsim 5 M_Z \sin\th_W$, this state is mostly bino, and therefore can
avoid lower bounds on neutralino masses, as we will now discuss.

\subsection{The light neutralino}\label{sec:bino}  % Matthew
The $R$-symmetry protects the, mostly bino, neutralino above from gaining a Majorana
mass within the globally supersymmetric model discussed so far.  However, we expect
that SUGRA effects will violate the global $R$-symmetry, and when this is combined
with SUSY breaking this could lead to an $R$-symmetry-violating Majorana mass for
the gauginos.  As $X$ has non-zero $R$-charge, gaugino masses of the form;
\begin{equation}
\int d^2 \theta \frac{\mb{X}}{M_P} \mb{W^\alpha W_\alpha} = \frac{F_X}{M_P} \widetilde W^\alpha \widetilde W_\alpha ~~,
\end{equation}
are forbidden.  This does not mean that Majorana masses are not generated, however, as there would likely exist anomaly-mediated contribution not greater than \cite{Randall:1998uk,Giudice:1998xp};
\begin{equation}
m_\lambda = \frac{\beta(g^2)}{2 g^2} m_{3/2} ~~,
\end{equation}
which, for the bino, implies a Majorana mass of
\begin{equation}
m_1 = \frac{11 \alpha}{4 \pi \cos^2 (\theta_W)} m_{3/2} = 8.9 \times 10^{-3}\, m_{3/2} ~~.
\end{equation}

Now, as we are considering low-scale mediation of SUSY-breaking, such as gauge mediation, then for squark and slepton masses at the TeV scale we require $F_X/M \lesssim 100$~TeV.  Furthermore, for the single-Higgs generation of down-type fermion masses we require $F_X/M^2 \gtrsim 1/35$.  Combining these relations we find that $F_X \lesssim 3.5 \times 10^{5} \,\text{TeV}^2$, and thus the gravitino mass is of order $m_{3/2} \lesssim 83$ eV.  Therefore we expect a Majorana bino mass of:
\begin{equation}
m_1 \lesssim 0.67 \text{ eV} ~~.
\end{equation}

Such a light neutralino, with SM couplings through sfermion-fermion-bino terms in the Lagrangian might make the reader uneasy.  However, it has recently been shown that very light neutralinos can be compatible with cosmology and collider constraints, so long as the neutralino is mostly bino \cite{Dreiner:2009ic}.

This can be understood quite simply.  Interactions between the large bino component of the neutralino and SM fermions proceed via sfermion exchange.  Thus if there exists a small hierarchy between soft scalar masses and the weak scale such that $\tilde{m} \gtrsim M_W$, then these interactions are suppressed in comparison with neutrino interactions, which proceed via electroweak boson exchange.

As the bino component carries no gauge charges, the only gauge interactions of the lightest neutralino arise due to its small Higgsino component.  Regarding electroweak interactions, the lightest neutralino behaves in a similar manner to a neutrino, however vertices involved in physical processes are suppressed by the square of the small bino-Higgsino mixing angle.  Thus it is clear that the lightest neutralino behaves very much like an additional neutrino, but with a suppressed SM-neutralino interaction strength.\footnote{Such a light neutralino cannot constitute the dark matter of the universe.  It is reasonable that the dark matter could be made up of axions \cite{Preskill:1982cy,Abbott:1982af,Dine:1982ah} or could originate from within some other hidden sector \cite{Steffen:2008qp,Hall:2009bx,Cheung:2010gj,Cheung:2010gk}.}

In order to demonstrate that such a particle is acceptable we summarize the results of a recent study on neutralino mass bounds in \aref{app:neut}, however for a thorough discussion we refer the reader to Ref. \cite{Dreiner:2009ic}.

\subsection{Avoiding a light neutralino}\label{sec:binomass}  % Matthew
Although a very light neutralino is compatible with current observations in particle physics, astrophysics and cosmology, it may be to the taste of some readers to remove any particles surplus to the SM with masses below the weak scale.  For the bino this can be achieved with a little additional model building.

The simplest solution is to introduce a gauge singlet chiral superfield, $\mb{S}$, with $R$-charge $R_{\mb{S}} = 0$.  In this way a TeV-scale Dirac bino mass can be generated through the operator:
\begin{equation}
\int\! d^2\th \frac{\mb{W'_\alpha}}{M} (\mb{S B^\a}) ~~.
\end{equation}
The problem that arises with the addition of this singlet is that, among other terms, a \kahler potential term $K \supset \mb{X ^\dagger X S}/M$ cannot be forbidden by the imposed symmetries.  This term leads to a large VEV for S, which depends on the SUSY breaking scale, and also the soft mass for $S$, as:
\begin{equation}
|\langle S \rangle| \sim \frac{F_X^2}{M} \frac{1}{\tilde{m}^2_s} ~.
\end{equation}

Thus this VEV potentially leads to a large $\mu$-term, hypercharge $D$-term, or even destabilization of the SUSY-breaking in the hidden sector.  This is a common affliction of models involving singlet scalar fields.  It is possible to build models of $R$-symmetric gauge mediation which avoid large tadpoles for $S$.  In \cite{Amigo:2008rc} a $C$-parity symmetry is imposed which forbids dangerous tadpole terms.  In \cite{Benakli:2010gi} particular structures in the couplings between the adjoints and the messengers are chosen and a large tadpole term is not generated.  Finally in \cite{Abel:2011dc} it is demonstrated that these tadpole terms can be avoided if the adjoint-messenger couplings respect $SU(5)$ or if the singlet originates from a complete $SU(5)$ adjoint multiplet.
\subsection{Neutrinos}\label{sec:neutrino}  % Rhys

Neutrino masses are straightforward to accommodate.  The only dimension five operator which
violates baryon or lepton number and is allowed by the symmetries is the Weinberg
operator $\mb{H^2 L^2}$, or more explicitly
\begin{equation}
    \frac{1}{M_*} \int\! d^2\th\, \e_{ab}\e_{cd}\mb{H}^a \mb{H}^c \mb{L}^b \mb{L}^d \supset
        \frac{1}{M_*} \int\! d^2\th\, (H^0)^2 (\nu_L)^2~,
\end{equation}
where $M_*$ is some mass scale.
This is exactly what we need to generate neutrino masses after electroweak symmetry
breaking.  The obvious way for this term to come about is from a K\"ahler potential term
\begin{equation}
    \int\!d^4\th\, \frac{\mb{X^\dagger}}{M^3} \mb{H^2 L^2}~,
\end{equation}
but fortunately this is forbidden by our $\IZ_2$ symmetry, as it would give rise to
Majorana masses of order $M_{\nu_L} \sim F_X v^2/M^3$, which are too large due to the
low scale of SUSY breaking.

We can in fact implement a standard seesaw mechanism.  We introduce singlet
chiral superfields $\mb{N}$ which are even under the $\IZ_2$ symmetry and have $R$-charge
$1$.  These can then be given large supersymmetric masses, but the fermions can also
get weak-scale Dirac masses with the left-handed neutrinos after electroweak symmetry
breaking.  The relevant Lagrangian is
\begin{equation}
    \cL_\nu = \int\!d^2\th \left( M_R^2 \mb{N}^2 + \l_\nu \mb{H L N} \right)~.
\end{equation}
The mass scale suppressing the Weinberg operator is therefore the Majorana mass of these
right-handed neutrinos, leading to acceptably small neutrino masses.\footnote{We make
the reasonable assumption that there are no $R$-charge 1 fields at the messenger scale
which have superpotential couplings to $\mb{HL}$.}

%%%
\section
[Flavour, CP and precision electroweak]
{\boldmath Flavour, $CP$ and precision electroweak}\label{sec:R-symmetry}  % Rhys and Matthew
As we will see, the SOHDM is surprisingly robust against constraints from FCNCs, $CP$-violation, and precision electroweak observables.
\subsection{Flavour-changing neutral currents}
The largest potential source for FCNCs lies in the sparticle spectrum.  Considering the flavour structure of \eqref{eq:yuk} and \eqref{eq:softlag}, we see that, whilst remaining consistent with any flavour symmetries, the sparticle soft masses can originate from terms of the form;
\begin{equation}\label{eq:soft}
    \begin{split}
        \cL_\mathrm{soft}  = & \int\!d^4\th \frac{\mb{X^\dagger X}}{M^2} \bigg(\mb{Q^\dagger} (\a_Q^2
            + a_{Q1} \lambda_{U} \lambda_{U}^\dagger + a_{Q2} \lambda_{D} \lambda_{D}^\dagger) \mb{Q} 
           + \mb{{U^c}^\dagger} (\a_U^2 + a_U \lambda_{U}^\dagger \lambda_{U} ) \mb{U^c} \\
           & + \mb{{D^c}^\dagger} (\a_D^2 + a_D \lambda_{D}^\dagger \lambda_{D} ) \mb{D^c}
           + \mb{{L}^\dagger} (\a_L^2 + a_L  \lambda_{E} \lambda_{E}^\dagger ) \mb{L} + \mb{{E^c}^\dagger} (\a_E^2
           + a_E \lambda_{E}^\dagger \lambda_{E} )\mb{E^c} \bigg) ~,
    \end{split}  
\end{equation}
where we are suppressing flavour indices, and neglecting higher powers of the Yukawa matrices $\lambda$.  It should be noted that in \eqref{eq:yuk} both $\lambda_{D}$ and $\lambda_{E}$ come dressed with the SUSY breaking field $X$, thus one would expect on general grounds to generate the additional non-diagonal terms in \eqref{eq:soft}, as well as the diagonal terms which may arise due to a low-scale mediation mechanism such as gauge mediation.  It is also important to recall that, as a result of \eqref{eq:yuk} the down-type quark Yukawas are of the form $(F_X/M^2) \lambda_D$, and similarly for the leptons.  This implies that if $F_X/M^2 \sim 1/35$ then the Yukawa matrix $\lambda_D$ can have large entries with $\lambda_D \sim \mathcal{O}(1)$ for the bottom quark entries.  Thus the non-diagonal components of \eqref{eq:soft} involving $\lambda_D$ are not necessarily small compared to the components involving $\lambda_U$.

We expect that FCNC processes within the SOHDM will be small and well within current bounds.  This is due to the following three effective FCNC-suppressing ingredients of this model:
\begin{itemize}
\item  Minimal flavour violation / flavour alignment.
\item  $R$-symmetry.  This forbids A-terms that lead to left-right sfermion mixing after
electroweak symmetry breaking, and also forbids Majorana gaugino masses.
\item  SM fermions couple at tree-level to a \emph{single} Higgs doublet.
\end{itemize}

It can be seen from \eqref{eq:soft} that, by extending the flavour rotations that diagonalise the SM fermion mass matrices to act on the whole supermultiplets, the resulting sfermion mass-squared matrices $\tilde{m}^2_{L},\tilde{m}^2_{E^c},\tilde{m}^2_{U^c}$ and $\tilde{m}^2_{D^c}$ will be diagonalised automatically.\footnote{As an example, we can see that if the quark rotation $\{d_R \rightarrow {U_R^d}^\dagger d_R, d_L \rightarrow U_L^d d_L\}$ diagonalises $\lambda_D$ then the same rotation for the squarks $\tilde{d}_R \rightarrow {U_R^d}^\dagger \tilde{d}_R$ diagonalises the $\tilde{d}_R$ mass-squared matrix.}  However, in order to diagonalise the mass-squared matrices $\tilde{m}^2_{Q_U}$ and $\tilde{m}^2_{Q_D}$ it will be necessary to perform a further rotation proportional to the CKM matrix, $V_{\text{CKM}}$.  This further rotation will introduce FCNC interactions at squark-quark-gluino and squark-quark-neutralino vertices.  However, the flavour structure at these vertices will be proportional to $V_{\text{CKM}}$ and will thus satisfy `Minimal Flavour Violation' (MFV) \cite{D'Ambrosio:2002ex}.  Thus within the SOHDM all FCNC processes are governed by the CKM matrix.  Furthermore, the only relevant operators in the effective Hamiltonian below the weak scale are those relevant in the SM.  This structure does not, however, require that the squark masses are degenerate.

It is possible that some other physics, beyond that described within this model, could lead to extra non-MFV terms in \eqref{eq:soft}.  However, the best-motivated source for such terms would be through gravity-mediation effects, which would be small ($\mathcal{O}(100 \text{eV})$) in the SOHDM as the scale of supersymmetry breaking is low.  The MFV assumption is therefore well-motivated within the current framework.

The $R$-symmetry plays a pivotal role in suppressing FCNC processes.  In \cite{Kribs:2007ac} it was shown that, when a supersymmetric model possesses an $R$-symmetry, the absence of Majorana gaugino masses and tri-linear A-terms, which generate left-right sfermion mixing, leads to a strong suppression of FCNC processes.  The suppression is effective enough that, with Dirac gaugino masses of order a few TeV, and vanishing left-right sfermion mixing, flavour violating sfermion masses (which violate MFV) of $\mathcal{O}(1)$ are allowed.

The detailed reasons for this extra suppression are described in \cite{Kribs:2007ac}, and we briefly summarise the results here.  For the case of $\Delta F = 2$ flavour violation the strongest constraints come from $K-\overline{K}$ mixing, and next strongest from B mixing.  The $R$-symmetry suppresses SUSY contributions to these processes as the usual troublesome dimension-five operators, such as;
\begin{equation}
\frac{1}{m_{\tilde{g}}} \tilde{d}^\ast_R \tilde{s}^\ast_L \overline{d}_R s_L ~,
\end{equation}
are forbidden by the $R$-symmetry, and the leading operators are dimension six.  In addition the Dirac gauginos lead to finite, rather than log-enhanced, radiative corrections to squark masses.  The result is that the box diagrams leading to $K-\overline{K}$ mixing are suppressed sufficiently to allow $\mathcal{O}(1)$ non-MFV flavour violation in the squark sector.  In \cite{Kribs:2007ac} it is shown that phases in squark masses are still constrained by limits on $\epsilon_K$.  In particular, with $\mathcal{O}(1)$ non-MFV flavour violation in the squark sector, the phases are constrained to be $\theta < 0.15$.  This constraint weakens if the first two generations of squarks are approximately degenerate, which is fortunately the case for the SOHDM if the dominant contributions to squark masses arise from flavour-diagonal gauge mediated terms.

For $\Delta F = 1$ flavour violation, such as $b \rightarrow s \gamma$, there is also a suppression due to the $R$-symmetry \cite{Kribs:2007ac}.  This is due to the fact that the Feynman diagrams for these processes involve a helicity flip, and this is not possible for an internal gaugino line as the opposite-helicity state has no tree-level couplings to SM fermions.  The only contributing diagrams then involve a helicity flip on an external line.  These contributions are sufficiently suppressed to allow  $\mathcal{O}(1)$ non-MFV flavour violation in the sfermion sector.  Similarly, constraints from $\epsilon'/\epsilon$ do not lead to strong constraints on flavour violation \cite{Kribs:2007ac}.

Thus the $R$-symmetry acts to efficiently suppress flavour-violating processes, in addition to the MFV in the SOHDM.\footnote{It has recently been noted that in R-symmetric models $\mathcal{O}(1)$ mixing in the slepton sector is not allowed, and the allowed mixing is $\lesssim \mathcal{O}(0.1)$.  This is due to limits on lepton flavour violation from processes such as $\mu \rightarrow e \gamma$ \cite{Fok:2010vk}.  As we have the additional assumption of MFV then slepton mixing parameters will be small and the SOHDM is safe from these constraints.}

\subsection
[CP-violation]
{$CP$-violation}\label{sec:CP}

We have seen above that flavour violation imposes no significant constraints on the
parameter space, but we must also consider new flavour-diagonal sources of
$CP$-violation.

First let us isolate physical phases in the Lagrangian.
We can choose the phases of $\mb{O}$ and $\mb{T}$ so that the Dirac mass parameters
$M_2$ and $M_3$ are real.  A rotation of $\mb{\eta}$ can then make $\l_T$ real.  Finally,
rephasing $\mb{H}$ can make $\mu$ real, thereby removing all supersymmetric phases in
our `flavourless' sector.  As a result, all new physical phases lie in the scalar soft masses.

At a typical point in MSSM parameter space, large electric dipole moments are generated
at one-loop for leptons and quarks, whereas experimentally, such dipole moments are
constrained to be small.  In the SOHDM, the $R$-symmetry forbids all such one-loop diagrams.

The other contribution to the electric dipole moment of the neutron is the dimension-six
three-gluon operator first discussed by Weinberg \cite{Weinberg:1989dx}.  This obtains
corrections from diagrams involving the Dirac gluinos, but it was argued in
\cite{Kribs:2007ac} that this imposes no strong constraints for TeV-scale masses.  

\subsection{Precision electroweak tests}

Since the SOHDM has a significantly modified electroweak sector, one might worry that
it is already ruled out by precision electroweak tests.  In particular, we can ask what
contributions are made to the parameters $S$ and $T$.

At tree level, there is a contribution to $T$ from the small VEV obtained by the triplet.
As before, we will write $v_T$ for this VEV; the tree level $\rho$ parameter is then
\begin{equation}
    \rho := \frac{M_W^2}{M_Z^2 \cos^2\th_W} = 1 + \frac{2 g^2 v_T^2}{M_Z^2 \cos^2\th_W}~.
\end{equation}
The experimental upper bound is roughly $\rho \lesssim 1.0012$ \cite{RPP}, which
implies $v_T \lesssim 3.6\GeV$.  The expression for the triplet VEV in the SOHDM was
given in \eref{eq:tripletVEV}.  As described in \sref{sec:model}, $\l_T$ is small, and if we take $\l_T \simeq 0$, we get
\begin{equation}
    v_T \simeq \left(\frac{g M_2 v}{8M_2^2 + 4 \tilde m_T^2}\right) v~,
\end{equation}
which for typical values of the parameters evaluates to $\lesssim 1\GeV$, so this gives no
significant constraints on the model.

In \cite{Martin:2010dc} the effects of an $SU(2)$ triplet superfield $\mb{T}$ on the $S$ and $T$ parameters at one-loop level are considered.  The trilinear coupling;
\begin{equation}\label{eq:trilinear}
    \cL_{\text{Tri}} = \int\!d^2\th\, \l_T \,\mb{H T \eta} ~,
\end{equation}
breaks the custodial symmetry of the Higgs sector, however, in \cite{Martin:2010dc} it is shown that even for large Yukawa couplings the $S$ and $T$ parameters lie within the current 68 \% confidence level ellipse.  In our model, this coupling is small ($\l_T \lesssim 1/35$) and one-loop corrections will remain within current bounds.

\section{Gauge Couplings}\label{sec:couplings}
Due to the addition of the $SU(3)$ octet superfield $\mb{O}$, the $SU(3)$ gauge coupling does not run at one-loop \cite{Martin:2010dc} at high scales.  In addition, above the mass of the octet, asymptotic freedom is lost at two-loops.  However, due to the two-loop suppression, the $SU(3)$ gauge coupling remains perturbative up to scales as high as $10^{18} \text{GeV}$ and no Landau pole problems arise.

It is also clear that, in comparison with the MSSM, there are no new contributions to the $U(1)_Y$ beta function, and $\mb{T}$ and $\mb{O}$ contribute differing amounts to the $SU(2)$ and $SU(3)$ beta functions, so that gauge coupling unification is lost.

One approach to recover unification is to add extra vector-like matter superfields $\mb{L',\overline{L}',E',\overline{E}',E',\overline{E}'}$, where the quantum numbers of the MSSM lepton doublet and right-handed electron are implied \cite{Fox:2002bu,Martin:2010dc}.  These fields can be given weak-scale vector-like masses, and dangerous mixings with SM fields can be forbidden with the imposition of an appropriate discrete symmetry.  If five singlets are also added then these new fields, in addition to the triplet and octet, would correspond to an adjoint representation of the GUT group $SU(3)_c \times SU(3)_L \times SU(3)_R \subset E_6$ \cite{Fox:2002bu,Martin:2010dc}.\footnote{One of these singlets would play the role of the adjoint $U(1)$ chiral superfield and three would take the place of right-handed neutrino superfields, leaving only one additional singlet superfield.}  Additional vector-like matter falling in a complete representation of this gauge group must then be added in order to implement gauge mediation.  This must contain at most two pairs of $\rep{3},\conjrep{3}$ under $SU(3)_c$ in order to maintain perturbativity up to the GUT scale.

An alternative approach based on $SU(5)$ would be to embed the triplet and octet in an adjoint of $SU(5)$ \cite{Fox:2002bu}.  This requires the addition of a singlet and the vector-like `bachelor' superfields $\mb{B,\overline{B}}$ with quantum numbers $(\rep{3},\rep{2}, -5/6)$ and $(\rep{3},\conjrep{2}, 5/6)$.  Again, these fields can be given vector-like masses and unification can be achieved \cite{Fox:2002bu}.  For a gauge-mediated scenario it would be appealing to use these bachelor fields as messengers.

\section{Collider Signatures}\label{sec:collider}
The SOHDM has a number of collider signatures which could be used to distinguish this model from the MSSM.  Some of these features arise due to the $R$-symmetry and are common in $R$-symmetric models, whereas others arise as a result of the single-Higgs nature of the model.

One striking aspect of $R$-symmetric models is the presence of Dirac, rather than Majorana, gluinos and neutralinos, with restricted production and decay channels.  The distinction between the Dirac and Majorana cases has been discussed in detail in \cite{Choi:2008pi,Choi:2009ue,Choi:2010gc}, and processes which are allowed within the MSSM, but are forbidden in an $R$-symmetric model, have been enumerated in \cite{Choi:2008pi}.  We summarise these processes below:
\begin{itemize}
\item Different-flavour quark-quark scattering:  $\sigma[q q' \to \{\tilde{q}_L \tilde{q}'_L,\tilde{q}_R \tilde{q}'_R\}]  = 0$
\item Different-flavour quark-antiquark scattering:  $\sigma[q \overline{q}' \to \{ \tilde{q}_L \tilde{q}'_R, \tilde{q}_R \tilde{q}'_L\}] = 0$
\item Squark-gluino production:  $\sigma[q g \to \{ \tilde{q}_L \tilde{g}_D, \tilde{q}_R \tilde{g}_D\}] = 0$
\item Gluino pair production:  $\sigma[q \overline{q} \to \{ \tilde{g}_D \tilde{g}_D, \tilde{g}^c_D \tilde{g}^c_D\}] = \sigma[g g \to \{ \tilde{g}_D \tilde{g}_D, \tilde{g}^c_D \tilde{g}^c_D\}]  = 0$
\end{itemize}
Similar alterations to the electroweak sector occur.  In \cite{Choi:2008pi} it is described how the differences in decay processes could be used to determine sfermion handedness through like-sign and unlike-sign dilepton signals at the LHC.  If we ignore the small Yukawa interactions of the first two generations, and focus on the gauge interactions, then right-handed sfermions do not couple at tree-level charginos, and thus decay dominantly to a bino-fermion pair.  However, left-handed sfermions can decay to a chargino-fermion pair.  Using this fact right- and left-handed sfermions can be discriminated.  Combining this with the differences in sfermion production processes listed above it is in principle possible to separate a Dirac theory from a Majorana theory to a high level of statistical significance at the LHC \cite{Choi:2008pi}.

Colour octet and weak triplet scalars also have a distinctive collider phenomenology \cite{Choi:2008ub,Choi:2009ue,Choi:2010gc}.  However, as we require, and expect, that these extra scalars are heavy ($\gtrsim 1$ TeV) we will not consider their phenomenology here.

Another feature of the SOHDM are the R-charge $\cR_{\mb{\eta}} = 2$ scalars.  In the MRSSM \cite{Kribs:2007ac}, in addition to the standard Higgs doublets, with $\cR_{\mb{H_u}} = \cR_{\mb{H_d}} = 0$, two R-charge $\cR_{\mb{R_u}} = \cR_{\mb{R_d}} = 2$ doublets are required.  These extra particles have been found to have interesting collider signatures \cite{Choi:2010an}.  Although the particle content in the SOHDM is reduced, and only the two doublets with $\cR_{\mb{H}} = 0$ and $\cR_{\mb{\eta}} = 2$ are present, the main features of collider phenomenology for these sets of particles are similar; with a purely standard-model initial state any $R$-charged particles must be pair produced at colliders.  For the $\cR_{\mb{\eta}} = 2$ doublet this occurs dominantly at the LHC through Drell-Yan production mediated by electroweak gauge bosons \cite{Choi:2010an}, and for masses below $250$ GeV the cross-section is $\mathcal{O}(10)$ fb.  Since $|\cR_{\mb{\eta}}| = 2$, in contrast to all other R-charged particles in the model, $\eta$-boson decay must result in a pair of light neutralinos, and four light neutralinos for any event involving the pair production of $\eta$-bosons.  

The SOHDM could also be discriminated from the MSSM or the MRSSM at the LHC through the observation of particles originating from the Higgs and electroweak gauge sectors.  We summarize the multiplicity of these particles in \tref{tab:weakpart}, where we exclude the adjoint scalars as we expect their masses to lie well above the TeV scale.  One can see that, in particular, the charged particle multiplicities differ for all three models, and this could be used to distinguish between these models at the LHC.

\begin{table}[t]
\begin{center}
%% The following adds some white space between row-dividing lines and text %%
\def\str{\vrule height12pt width0pt depth7pt}
\begin{tabular}{| c | c | c | c | c |}
    \hline\str
    ~Model ~&~ Scalar ~&~ Fermion ~ \\
    \hline\str
    ~~MSSM & ~~$3^0_R, 2^\pm$ & $4^0_M, 2^\pm$ \\
    \hline\str
    ~~MRSSM & ~~$3^0_R, 2^0_C, 4^\pm$ & $4^0_D, 4^\pm$ \\
    \hline\str
    ~~SOHDM & ~~$1^0_R, 1^0_C, 2^\pm $ & $2^0_D,1^0_M, 3^\pm$ \\
    \hline
\end{tabular}
\parbox{.95\textwidth}
{\caption{\label{tab:weakpart}
\small Multiplicity of electroweak-charged particles originating from the Higgs and gauge sectors of related supersymmetric models.  We only include particles with masses at the TeV scale, i.e.\ omitting adjoint scalars.  Neutral and charged particles are discriminated by the superscript $(0,\pm)$.  For neutral scalars the subscript denotes whether the scalar is real (1 d.o.f.) or complex (2 d.o.f.), and for neutral fermions the subscript denotes whether the fermion is Majorana or Dirac.}}
\end{center}
\end{table}

Non-degenerate and non-diagonal flavour structure in sfermion masses would also hint at some underlying structure which protects the extra particles from generating unacceptable FCNCs in the SM sector, providing indirect evidence for the existence of a suitable extended $R$-symmetry.

At future colliders, precision Higgs physics could provide strong support for a single Higgs-doublet model.  This is because the couplings of the Higgs to SM fermions would be the same as in the SM, which is not the case in commonly considered supersymmetric models.  Finally, it is conceivable that a super-LHC (or `SSC') running at $\sqrt{s} \sim 100$ TeV could uncover the messenger and SUSY-breaking sectors as, in the SOHDM, both sectors are required to exist at this scale.

\section{Conclusions}\label{sec:con}
We have described a supersymmetric model in which only a single Higgs doublet participates in electroweak symmetry breaking, this doublet providing mass to all SM fermions.  The extra doublet superfield required for anomaly cancellation, $\mb{\eta}$, is merely a `spectator' field, and does not acquire a VEV.  The model has a number of distinctive features, such as an anomaly-free $R$-symmetry which is imposed to protect $\eta$ from mixing with $H$, but also helps to ameliorate FCNC problems of supersymmetric models.  This is particularly advantageous in models where fermion mass generation is tied to SUSY-breaking.  Thus a non-degenerate or non-diagonal sfermion-mass flavour structure is allowed.  The model \emph{requires} a low scale of SUSY-breaking, and the simplest version predicts a very light neutralino with mass $\mathcal{O}(0.7)$ eV.

The many attractive features of the SOHDM mean its implications for the LHC
need to be seriously considered.  It also motivates further exploration of models of
low-scale mediation of SUSY-breaking which not only maintain an
$R$-symmetry, but also generate the flavour structure of the down-type fermions as in
\eqref{eq:yuk}.

\acknowledgments
We gratefully thank Stephen West for stimulating discussions.  MM is supported by an STFC Postgraduate Studentship.  R.D. is supported by the Engineering and Physical Sciences
Research Council [grant number EP/H02672X/1].  JMR and MM also acknowledge support by the EU Marie Curie Network �UniverseNet� (HPRN-CT-2006-035863), and JMR by a Royal Society Wolfson Merit Award. 
\vspace{-10pt}

\appendix
\section{Bounds on a light Neutralino}\label{app:neut}
\vspace{-5pt}

\subsection{Collider bounds}\label{binocoll}  % Matthew
The most stringent collider bounds come from LEP.  As LEP operated at $\sqrt{s} \le 208$ GeV, associated production of the lightest neutralino via $e^+ e^- \rightarrow \widetilde{\chi}_1^0 \widetilde{\chi}_2^0$ would have been kinematically forbidden for $\{ \mu, M_2\} \gtrsim 250$ GeV, due to the large mass of $\widetilde{\chi}_2^0$.  Thus, although such a channel would give clear signals following the decay  of $\widetilde{\chi}_2^0$, it does not necessarily constrain a very light neutralino.

Radiative neutralino production via $e^+ e^- \rightarrow \widetilde{\chi}_1^0 \widetilde{\chi}_1^0 \gamma$ provides another potential search channel.  However, for a mostly bino $\widetilde{\chi}_1^0$, radiative neutrino production, $e^+ e^- \rightarrow \overline{\nu} \nu \gamma$, generates a large background to this process.
\vspace{-5pt}

\subsection{Precision electroweak}\label{binoprec}  % Matthew
As the decay channel $Z_0 \rightarrow \widetilde{\chi}_1^0 \widetilde{\chi}_1^0$ is kinematically allowed, precision measurements of the total and invisible $Z_0$-width, $\Gamma_Z$ and $\Gamma_{\text{inv}}$, are sensitive to a very light neutralino.  In the minimal version of the SOHDM the decay $Z_0 \rightarrow \widetilde{\chi}_1^0 \widetilde{\chi}_1^0$ can proceed at tree-level due to the small Higgsino component of $\widetilde{\chi}_1^0$.  In addition the $Z_0$ can decay to the large bino components of $\widetilde{\chi}_1^0$ through loops involving sfermions.  Thus, by decreasing the Higgsino fraction of $\widetilde{\chi}_1^0$, and increasing the sfermion masses, these processes can be suppressed compared to decays to neutrinos.

By using the results of \cite{Heinemeyer:2007bw}, in which the processes $Z_0 \rightarrow \widetilde{\chi}_1^0 \widetilde{\chi}_1^0$ and $Z_0 \rightarrow \overline{f} f$ have been calculated at $\mathcal{O}(\alpha)$ within the MSSM, the $Z_0$-width was studied for a massless neutralino in \cite{Dreiner:2009ic}.  In this study the bino Majorana mass $M_1$ was set such that the lightest neutralino is massless, and one particular choice of parameters was with sfermion masses $M_{SUSY} = 600$ GeV, $\tan{\beta} = 10$, and $A_{\tau} = A_t = A_b = M_{\tilde{g}} = M_A = 600$ GeV.  Upon calculating contributions to the $Z_0$-width it was found that, within the MSSM, the Z-width observable $\Gamma_Z$ was within $1 \sigma$ of the experimental value for $|\mu| \gtrsim 200$ GeV and $M_2 \gtrsim 100$ GeV, and this small discrepancy reduced rapidly for larger values of $\mu$.

The invisible width $\Gamma_{\text{inv}}$ was within $2 \sigma$ of the experimental value for $|\mu| \gtrsim 200$ GeV and $M_2 \gtrsim 100$ GeV, however there was a $1 \sigma$ deviation across the entire $\mu-M_2$ plane.  This should come as no surprise however, as the SM prediction for $\Gamma_{\text{inv}}$ is $1.8 \sigma$ greater than the experimental value \cite{:2005ema} thus the comparative decrease in quality of fit with the addition of a light neutralino is small.

The study \cite{Dreiner:2009ic} therefore shows that $\Gamma_Z$ and $\Gamma_{\text{inv}}$ cannot exclude a massless neutralino within the MSSM.  Although the SOHDM is in some ways quite distinct from the MSSM, the important features of this study are common to both scenarios.  In particular, in both cases the light neutralino is mostly bino, with a small ($< 20 \%$) Higgsino component.  Thus we find it reasonable to conclude that $Z_0$-width measurements do not exclude the very light neutralino in the minimal version of the SOHDM.

In \cite{Dreiner:2009ic} a similar analysis of the impact of light neutralinos on $M_W$ and $\sin^2 \theta_{\text{eff}}$ within the MSSM was performed.  For two selected sets of soft SUSY-breaking parameters, detailed in \cite{Dreiner:2009ic}, it was show that $M_W$ lies within the experimental $1 \sigma$ boundary, as does $\Gamma_Z$ again.  It is interesting that $\sin^2 \theta_{\text{eff}}$ lies outside the experimental $1 \sigma$ boundary, however, as detailed in \cite{Heinemeyer:2007bw}, raising the soft scalar masses and/or the $\mu$ parameter improves this fit, and can lead to agreement at the $1 \sigma$ level whenever $\tilde{m} \gtrsim 700$ GeV.

Electric dipole moments and the anomalous magnetic moment of the muon were also considered for the MSSM with a light neutralino in \cite{Dreiner:2009ic}, where it was found that the SUSY contributions to EDMs go to zero for a massless $\widetilde\chi_1^0$.  Further, the variation of $(g-2)_\mu$ stays well below the current experimental uncertainty.  Thus no lower limit on the mass of the lightest neutralino in the MSSM can be set with these measurements.
\vspace{-5pt}

\subsection{Rare meson decays}\label{binodec}  % Matthew
In \cite{Dreiner:2009ic} decays of both pseudoscalar and vector mesons to bino pairs are considered within the MSSM.  These decays involve loops containing sfermions, and it is found that for sfermion masses of $\tilde{m} = 300$ GeV the branching ratios for the decays $\{\pi_0, \eta, \eta', B_s\} \rightarrow \widetilde{B} \widetilde{B}$, $\{\phi, J/\psi, \Upsilon(1S), \rho, \omega\} \rightarrow \widetilde{B} \widetilde{B}$, and $K^+ \rightarrow \pi^+ \widetilde{B} \widetilde{B}$ all lie well below current bounds.
\vspace{-5pt}

\subsection{Astrophysical bounds}\label{binoastro}  % Matthew
Neutrinos are produced in large abundance during a supernova explosion through electron-positron annihilation and neutrino-strahlung in nucleon scattering.  As their mean free path is smaller than the supernova core size these neutrinos slowly diffuse out.  After $\mathcal{O}(10 \text{ sec})$ the temperature-dependent mean free path exceeds the core size and they escape.  During Supernova 1987a such neutrinos were observed.

A similar process can occur for very light neutralinos, whereby they are produced in a supernova through similar processes to neutrinos.  If they interact more weakly than neutrinos, and have a mean free path exceeding the core size, they could escape the supernova and carry away a large amount of energy.  If this occurs and the energy loss is too great the neutrino mean free path will increase more rapidly, reducing the timescale over which neutrinos diffuse out.

In \cite{Dreiner:2009ic} it was shown that for the case of very weakly interacting neutralinos, in order that neutralino radiation energy losses do not exceed the Raffelt bound of $\leq 10^{52}$ erg, it is necessary that the production processes are sufficiently suppressed.  This leads to the requirement that the selectron mass exceeds $m_{\tilde{e}} \geq 1.2$ TeV, since the process $e^+ e^- \rightarrow \tilde{\chi}_1^0 \tilde{\chi}_1^0$ is suppressed by four powers of the selectron mass.  Similar, although much less restrictive bounds hold for the squark masses, requiring $m_{\tilde{q}} \geq 360$ GeV.
Alternatively, if the neutralinos have stronger interactions with SM particles it is possible that their mean free path is less than the core size.  In this case, the neutralinos will slowly diffuse out and the energy loss due to neutralino radiation will be suppressed while they are trapped within the core of the supernova.  It was found in \cite{Dreiner:2009ic} that in this scenario squark and slepton masses of $m_{\tilde{e}}, m_{\tilde{q}} < 300$ GeV are compatible with observations.

It would seem from this discussion that selectron masses in the range $300 < m_{\tilde{e}} \leq 1200$ GeV are excluded, however it should be noted that to-date successful simulations of a full supernova explosion have not yet been performed, and thus the above excluded regions may be subject to change in the future.

\subsection{Cosmological bounds}\label{binocos}  % Matthew
A very light neutralino, which is relativistic at freeze-out, will contribute to hot dark matter.  However it is known that early universe cosmology is most compatible with cold, non-relativistic, dark matter.

In order that the light neutralino does not constitute too much hot dark matter, and does not suppress structure formation, it is necessary that the energy density of hot dark matter is consistent with observations, and the Cowsik-McClelland bound is satisfied.
One can see how this arises, as the energy density of a neutrino is proportional to its mass.  Therefore if we reduce the mass of the heaviest neutrino this leaves room for energy density due to another hot dark matter component.  Calculation of the neutralino relic density leads to the requirement that:
\begin{equation}
m_{\widetilde{\chi}_1^0} \lesssim 0.7 \text{ eV} ~~.
\end{equation}
Thus a very light, mostly bino, neutralino is consistent with structure formation.  It is intriguing that this upper limit on the lightest neutralino mass lies just above the Majorana bino mass we expect to be generated due to anomaly-mediation.

Thus we see, thanks to the study \cite{Dreiner:2009ic}, that a mostly bino neutralino, with mass $m_{\widetilde{\chi}_1^0} \lesssim 0.7$ eV, is consistent with direct production collider bounds, precision electroweak observables, branching ratios in rare meson decays, supernova cooling rates, and structure formation in the early universe.  Thus the prediction of a very light, mostly bino, neutralino in the minimal version of the SOHDM is phenomenologically acceptable.
\vspace{-10pt}

\bibliographystyle{JHEP}
\bibliography{SingleHiggsrefs}

\end{document}